\documentclass{jnmp}
\JNMPnumberwithin{equation}{section}

\usepackage{amsmath}

\def\ord{\mathop{\rm ord}\nolimits}

\newtheorem{theorem}{Theorem}

\begin{document}
\renewcommand{\evenhead}{A Sergyeyev and J A Sanders}
\renewcommand{\oddhead}{Nonlocal Symmetries for the CDIS Equations}
\thispagestyle{empty}

\FirstPageHead{10}{1}{2003}{\pageref{artur-firstpage}--\pageref{artur-lastpage}}{Article}

\copyrightnote{2003}{A Sergyeyev and J A Sanders}

\Name{A Remark on Nonlocal Symmetries for
the Calogero--Degasperis--Ibragimov--Shabat
Equation}\label{artur-firstpage}

\Author{Artur SERGYEYEV~$^\dag$ and Jan A SANDERS~$^\ddag$}

\Address{$^\dag$~Silesian University in Opava, Mathematical
Institute,\\
~~Bezru\v{c}ovo n\'am. 13, 746~01 Opava, Czech Republic\\
~~E-mail: Artur.Sergyeyev@math.slu.cz\\[10pt]
$^\ddag$~Vrije Universiteit, Faculty of Science,
Division of Mathematics and Computer Science,\\
~~De Boelelaan 1081a, 1081 HV Amsterdam, The Netherlands\\
~~E-mail: jansa@cs.vu.nl}

\Date{Received May 02, 2002; Revised June 23, 2002;
Accepted June 24, 2002}

\begin{abstract}
\noindent
We consider the Calogero--Degasperis--Ibragimov--Shabat (CDIS)
equation and find the complete set of its nonlocal symmetries
depending on the local variables and on the integral of the only
local conserved density of the equation in question. The
Lie algebra of these symmetries turns out to be a central
extension of that of local generalized symmetries.
\end{abstract}

\section{Introduction}
The existence of infinite-dimensional Lie algebra of commuting
higher order symmetries for a system of PDEs is well known to be one of
the most important signs of its integrabi\-li\-ty, see e.g.\
\cite{artur:blasz, artur:olv_eng2, artur:sh_s, artur:sok}.
This algebra can be extended, see e.g.\ \cite{artur:blasz}, to a noncommutative
algebra (which is
often
referred as a {\em hereditary algebra}, see
\cite{artur:blasz} and references therein) of
polynomial-in-time (and possibly nonlocal)
symmetries. In (2+1) dimensions these symmetries are
polynomials in time $t$ of arbitrarily high degree, while in (1+1)
dimensions one usually can construct only the symmetries
which are at most linear in time~\cite{artur:blasz}.

For a long time the only known
(1+1)-dimensional nonlinear evolution
equation possessing
symmetries
being polynomials in time of arbitrarily high degree
was
the Burgers equation,
see \cite{artur:vk}
for the complete description of its symmetry algebra. It is natural to ask whether
there exist other (1+1)-dimensional evolution equations having the same property.
In \cite{snmp} we have answered this question in affirmative and shown
that the Calogero--Degasperis--Ibragimov--Shabat
(CDIS) \cite{artur:cd, artur:cal, artur:is, artur:sw-rt, artur:sh_s, artur:ss}
equation also possesses a heredi\-tary algebra of polynomial-in-time
symmetries
which, exactly as in the case of Burgers equation, is its complete symmetry algebra
in the class of {\em local} higher order symmetries.

The CDIS equation (\ref{artur:is}) has only one nontrivial
local conserved density $\rho=u^2$, so the natural next step in
analyzing this equation is to
consider its symmetries involving a nonlocal variable $\omega$ being the
integral of this density. Theorem 1 below provides the complete
characterization of symmetries which depend on this variable and on a
finite number of local variables. It turns out that, unlike the local case,
the Lie algebra of these symmetries possesses a nontrivial one-dimensional
center, spanned by (the only) genuinely nonlocal symmetry.

\section{Symmetries of the CDIS equation}

The Calogero--Degasperis--Ibragimov--Shabat (CDIS) equation
has the form \cite{artur:cd, artur:is}
\begin{equation}\label{artur:is}
u_{t}=u_{3}+3 u^{2} u_{2}+9 u u_{1}^{2}+ 3 u^{4} u_{1}\equiv F,
\end{equation}
Here $u_{j}=\partial^{j} u/\partial x^{j}$ and $u_0\equiv u$; see \cite{snmp} for
the further details on notation used.

Let us mention that (\ref{artur:is}) is the only equation among third order
(1+1)-dimensional scalar polynomial $\lambda$-ho\-mo\-ge\-ne\-ous
evolution equations with $\lambda=1/2$ that possesses
infinitely many $x,t$-independent local generalized symmetries~\cite{artur:sw}.

Consider a nonlocal variable $\omega$ defined (cf.\ e.g.\ \cite{artur:ss,
artur:vin,artur:vk}) by the relations
\begin{equation}\label{om}
\p\omega/\p x=u^2,\qquad \p\omega/\p t=2u u_2 + 6 u^3 u_1+u^6-u_1^2.
\end{equation}
Note that the CDIS equation is linearized into $v_t=v_3$ upon setting
$v=\exp(\omega) u$ \cite{artur:sh_s}.

The quantity $\rho=u^2$ is \cite{artur:sh_s, artur:ss}
the only nontrivial local
conserved density
for (\ref{artur:is}),
but (\ref{artur:is}) has \cite{artur:ss} a Noether operator, i.e., an operator that
sends conserved covariants to symmetries, see
e.g.\
\cite{artur:ff1} for more details on such operators, of the form
$\exp(-2\omega)$ and
infinitely many nontrivial conserved densities explicitly dependent on
$\omega$ \cite{artur:ss}.

We shall call a function $G(x,t,\omega,u,u_1,\dots,u_k)$ a {\em symmetry}
of CDIS equation, if
\begin{equation}\label{sym}
D_{t}(G)-F_{*}(G)=0,
\end{equation}
where $F_{*}=\sum\limits
_{i=0}^{3}\p F/\p u_i D^i$, and
$D\equiv D_x=\p/\p x+u^2 \p/\p\omega+\sum\limits_{i=0}^{\infty} u_{i+1}
\partial/\partial u_{i}$ and $D_t=\p/\p t+\left(2u u_2 + 6 u^3
u_1+u^6-u_1^2\right)\p/\p\omega+
\sum\limits_{i=0}^{\infty} D^{i}(F)\partial/\partial
u_{i}$ are the operators of total $x$- and $t$-derivatives.
Note that our definition of nonlocal symmetries
is a particular case of
the usual one, cf.\ e.g.\ \cite{artur:blasz}, but in terminology of
\cite{artur:vin} the solutions of (\ref{sym}) are referred as {\it
shadows} of symmetries.

For any function $H=H(x,t,\omega,u,u_1,\dots,u_q)$ we define its
{\em order} $\ord H$ as a greatest integer $m$ such that $\p
H/\p u_m\neq 0$, and set
\[
H_{*}=2\p H/\p\omega D^{-1}\circ u+
\sum\limits_{i=0}^{\ord H}\p H/\p u_i
D^i.
\]
Here $\circ$ denotes a composition law induced by
`generalized Leibnitz rule'
(see e.g.\ \cite{artur:mik1, artur:sh_s})
\[
D^{k}\circ f=\sum\limits_{j=0}^{\infty}
\frac{k(k-1)\cdots(k-j+1)}{j!}D^{j}(f)D^{k-j}.
\]
A function $H=H(x,t,\omega,u,u_1,\dots,u_q)$ is called {\em local} if $\p
H/\p\omega=0$.

Let $S_{\mathrm{CDIS}}^{(k)}$ be the set of
symmetries of the form
$G(x,t,\omega,u,u_1,\dots,u_k)$
for the CDIS
equation,
and let
$S_{\mathrm{CDIS}}=\bigcup_{j=0}^{\infty}
S_{\mathrm{CDIS}}^{(j)}$,
$S_{\mathrm{CDIS},j}=
S_{\mathrm{CDIS}}^{(j)}/S_{\mathrm{CDIS}}^{(j-1)}$
for $j\geq 1$, and
$S_{\mathrm{CDIS},0}=S_{\mathrm{CDIS}}^{(0)}/\Theta_{\mathrm{CDIS}}$,
where $\Theta_{\mathrm{CDIS}}=\{G(x,t,\omega)|G\in
S_{\mathrm{CDIS}}\}$.

Suppose that $k\equiv \ord G\geq 1$. Then differentiating
the left-hand side of (\ref{sym}) with respect to
$u_{k+2}$ and equating the result to zero yields $D(\p G/\p u_k)=0$.
Hence, in analogy with Section~5.1 of \cite{artur:olv_eng2},
for any symmetry $G\in S_{\mathrm{CDIS}}$ of order
$k\geq 1$
we have
\begin{equation} \label{artur:lead_artur}
  \partial G/\partial u_{k} = c_{k}(t),
\end{equation}
where $c_{k}(t)$ is a function of $t$.

In particular,
any symmetry $G\in S_{\mathrm{CDIS}}^{(2)}$
is of the form $G=c(t)u_2+g(x,t,\omega,u,u_1)$.
Substituting this expression into (\ref{sym}),
collecting the coefficients at $u_4$, $u_3$, $u_2$,
equating to zero these coefficients and the sum of remaining terms on
the left-hand side of (\ref{sym}),
and solving the resulting system of equations for $c(t)$ and
$g$ readily shows that  $c(t)=0$ and
$G$ is a linear combination (with constant coefficients) of $u_1$ and
$W=\exp(-2\omega)u$.

Below we assume
without loss of generality that
any symmetry $G \in S_{{\rm CDIS},k}$, $k \geq 1$, vanishes if
the relevant function $c_{k}(t)$ is identically equal to zero.

Further differentiating (\ref{sym}) with respect to
$u_{k+1}$ and  $u_k$ and then with respect to $x$ shows that,
exactly as in \cite{artur:sergy-romp},
for $k\geq 2$ we have $\p^2 G/\p u_{k-1}\p x=0$, and for $k\geq 3$
\begin{equation}\label{difx}
\partial^2 G/\partial x\partial u_{k-2}=\dot c_{k}(t)/3.
\end{equation}

Taking into account that $G\in S_{\rm CDIS}$ implies $\tilde
G=\partial^r G/\partial x^r \in S_{\rm CDIS}$, and successively
using (\ref{difx}), we find that for $k\geq 2$ and $r\leq [k/2]-1$
we have $\mathop{\rm ord}\nolimits\tilde G\leq k-2r$ and
\begin{equation}\label{difx2}
\partial \tilde G/\partial u_{k-2r}=(1/3)^r d^r c_{k}(t)/d t^r.
\end{equation}
For $r=[k/2]-1$ we have
$\mathop{\rm ord}\nolimits\tilde G\leq 3$. The straightforward
computation shows that all symmetries from
$S_{\rm CDIS}^{(3)}$ are at most linear in $t$, and thus
the function
$c_{k}(t)$ satisfies the equation $d^m c_{k}(t)
/d t^m=0$ for $m=[k/2]+1$. Hence, $\dim S_{{\rm CDIS},k}\leq
[k/2]+1$ for $k\geq 1$, i.e., the dimension of the quotient
space of symmetries of the form
$G=G(x,t,\omega,u,u_1,\dots,u_k)$ modulo the space of symmetries
of the form $G=G(x,t,\omega,u,u_1,\dots,u_{k-1})$ does not exceed
$[k/2]+1$ for $k\geq 1$. From this it
is immediate that all odd-order symmetries from $S_{\mathrm{CDIS}}$
are exhausted by local ones,
as we can exhibit exactly $[k/2]+1$ such symmetries
of order $k$ for each odd~$k$~\cite{snmp}.
\looseness=-1

Furthermore, as all symmetries of the form
$G(x,t,\omega,u,u_1,u_2)$ are exhausted by $u_1$ and $W$, in analogy with
Theorem 2 of \cite{artur:sergy-romp} we can show that all
symmetries of the form
$G(x,t,\omega,u,u_1,\dots,u_k)$ of the CDIS equation
are polynomial in time $t$ for all
$k\in\mathbb{N}$. Indeed, assume this result to be proved
for the symmetries of order $k-1$ and let us prove it for
symmetries of order $k$.
It readily follows from the above
that the function $c_k(t)=\p G/\p u_k$ is a polynomial in $t$
of degree not higher than $[k/2]$. Therefore, $\p^m G/\p t^m$,
where $m=[k/2]+1$, is a symmetry of CDIS of order not higher
then $k-1$ and thus is polynomial in $t$ by assumption,
whence we readily see that $G$ is polynomial in $t$ as well.
The induction on $k$, starting from $k=2$, completes the
proof.

Now let us turn to the study of time-independent symmetries
of the CDIS equation. This equation is well known to have
infinitely many
$x,t$-independent local generalized symmetries, hence it has
\cite{artur:ibrbook} a formal symmetry of infinite rank of
the form
$\mathfrak{L}=D+\sum
_{j=0}^{\infty}a_{j}D^{-j}$,
where $a_j$
are some $x,t$-independent local functions.

Taking the directional derivative of (\ref{sym}), we
find  that for any symmetry
$G\in S_{\mathrm{CDIS}}$ of order $k$ the quantity
$G_*$
is a formal symmetry of
rank not lower than
$k+1$ for the CDIS equation, and therefore (cf.\ e.g.\
\cite{artur:mik1, artur:sh_s}), provided $k\geq 1$ and $\p
G/\p t=0$, we have
\[
G_*
=\sum\limits_{j=1}^{k}c_j\mathfrak{L}^j+\mathfrak{B},
\]
where $c_j$ are some constants and
$\mathfrak{B}=\sum\limits
_{j=-\infty}^{0}b_j D^j$,
$b_j$ are some $t$-independent local functions.
\looseness=-1

From this equation we infer that (cf.\ \cite{snmp})
any
symmetry
$G(x,\omega,u,u_1,\dots,u_k)\in S_{\mathrm{CDIS}}$, $\ord G\geq 1$,
can be written as
\begin{equation}\label{artur:repst}
G=G_{0}(u,u_1,\dots,u_{k})+Y(x,u,\omega).
\end{equation}

It can be easily seen that $\partial Y/\partial x=\partial
G/\partial x$ and
$\partial Y/\partial\omega=\partial G/\partial\omega$
are time-independent symmetries of the CDIS equation of order not higher
than zero.
We readily conclude from the above that
$\partial Y/\partial x=c_1 W$ and $\partial Y/\partial \omega=c_2 W$
for some constants $c_1$, $c_2$.
As $\partial^2 Y/\partial x\partial\omega
=\partial^2 Y/\partial \omega\partial x$, we find that
$c_1=0$, and thus $\partial Y/\partial x=0$, so any
time-independent symmetry $G$ of
order $k\geq 1$ for the CDIS equation is $x$-independent as well, and
$G=G_{0}(u,u_1,\dots,u_{k})+ c W$
for some constant $c$.
As we have already shown above, the only symmetry of order zero or less from
$S_{\mathrm{CDIS}}$ is $W$. Thus, we conclude that any time-independent symmetry of
the form
$G(x,\omega,u,u_1,\dots,u_k)$ is $x$-independent as well.

Using the symbolic method, it can
be shown \cite{artur:sw} that the CDIS equation has no even order
$t,x$-independent
local generalized symmetries, so its only even order
time-independent symmetry (in the class of symmetries of the form
$G(x,\omega,u,u_1,\dots,u_k)$) is
$W$.

Now let us show
that the same result holds true for time-dependent symmetries
as well. The CDIS equation is
invariant under the scaling
symmetry $K=3 t F+  x u_1+ u/2$.
Therefore, if a symmetry
$Q$ contains the terms of weight
$\gamma$ (with respect to the weighting induced by $K$,
cf.\ \cite{artur:bsw}, when the weight of $u$ is $1/2$, the weight of
$\omega$ is $0$, the weight of
$t$ is $-3$, the weight of $x$ is $-1$,
and the weight of $u_j=j+1/2$), there
exists a
homogeneous symmetry
$\tilde Q$ of the same weight $\gamma$.
We shall write this as $\mathop{\rm wt}\nolimits(\tilde Q)=\gamma$.
Note that
$[K,\tilde Q]=(\gamma-1/2)\tilde Q$.

If $G \in S_{{\rm CDIS}, k}$, $k \geq 1$, is a polynomial in $t$
of degree $m$, then its leading coefficient
$\partial G/\partial u_k=c_{k}(t)$ also is
a polynomial in $t$ of degree $m'\leq m$,
i.e., $c_{k}(t)=\sum\limits_{j=0}^{m'}t^{j} c_{k,j}$, where $c_{k, m'} \neq 0$.
Consider
$\smash{\tilde G= \partial^{m'} G/\partial t^{m'}}\in S_{\rm CDIS}^{(k)}$.
We have
$\partial\tilde G/\partial u_k={\rm const}\neq 0$,
hence $\tilde{G}$ contains the terms of the weight $k+1/2$. Let $P$ be
the sum of all terms of weight
$k+1/2$ in $\tilde G$. Clearly, $P$ is a homogeneous symmetry of
weight $k+1/2$ by
construction, $\mathop{\rm ord}\nolimits
P=k$ and $\partial P/\partial u_k$ is a nonzero constant.
Next, $\partial P/\partial t\in S_{{\rm CDIS}}$
is a homogeneous symmetry
of weight $k+7/2$.
Obviously,  $\mathop{\rm ord}\nolimits \partial P/\partial t\leq k-1$.
By the above, all symmetries in
$S_{{\rm CDIS}}$ are polynomial in $t$, and thus for any
homogeneous $B\in S_{{\rm CDIS}}$,
$b\equiv\mathop{\rm ord}\nolimits B\geq 1$, we have
$\partial B/\partial u_b=t^r c_b$, $c_b={\rm const}$
for some $r\geq 0$. Hence,
$\mathop{\rm wt}\nolimits(B)=b-3r+1/2\leq b+1/2$,
and for $k\geq 1$ the set $S_{{\rm
CDIS}}$ does not contain
homogeneous symmetries $B$ such
that $\mathop{\rm wt}\nolimits(B)=k+7/2$ and
$\mathop{\rm ord}\nolimits B\leq k-1$, so $\partial P/\partial t=0$.

Thus, we conclude that
the existence of a time-independent symmetry
of order $k\geq 1$ from $S_{{\rm CDIS}}$
is a necessary condition for the existence of
a polynomial-in-time symmetry $G\in
S_{{\rm CDIS}}$ of the same order $k$.
Moreover, by the above all symmetries from
$S_{{\rm CDIS}}$ are polynomial in $t$. Hence, the fact that the
CDIS equation has no
time-independent symmetries $G(x,\omega,u,u_1,\dots,u_k)$ of even order
$k\geq 2$ immediately implies the absence
of any {\it time-dependent}
symmetries of even order $k\geq 2$ belonging to $S_{{\rm CDIS}}$.

Summing up the above results, we infer that the space $S_{\mathrm{CDIS}}$
is spanned by the symmetry $W=\exp(-2\omega)u$, and by {\em local}
generalized symmetries of the CDIS equation. The latter were found
in \cite{snmp}, and can be described in the following way.

Define the commutator of two functions $f$ and $g$ of
$x,t,\omega,u,u_1,\dots$ as (cf.\ e.g.\
\cite{artur:blasz, artur:mik1, artur:sh_s})
\[
[f,g]=g_{*}(f)-f_{*}(g).
\]

Set $\tau_{m,0}=x^{m} u_1 + m x^{m-1} u/2$, $m=0,1,2\dots$, and
$\tau_{1,1}=
x(u_{3}+3 u^{2} u_{2}+9 u u_{1}^{2}+ 3 u^{4} u_{1})$
$+\,3u_2/2+5u_1u^2+u^5/2$.
The latter is the first nontrivial master symmetry for the CDIS
equation,  see e.g.\ \cite{artur:sw, artur:sw-rt}.
The quantities $\tau_{0,0}$, $\tau_{1,0}$, $\tau_{2,0}$, and
$\tau_{1,1}$ meet the requirements of Theorem 3.18 from~\cite{artur:blasz}, whence
\begin{equation}
[\tau_{m,j},\tau_{m',j'}]=((2j'+1)m-(2j+1)m')
\tau_{m+m'-1,j+j'},\label{artur:comm}
\end{equation}
where $\tau_{m,j}$ with $j>0$ are defined inductively by
means of (\ref{artur:comm}), i.e.,
$\tau_{0,j+1}=\frac{1}{2j+1}[\tau_{1,1},\tau_{0,j}]$,
$\tau_{m+1,j}=\frac{1}{2+4j-m}[\tau_{2,0},\tau_{m,j}]$, see
\cite{artur:blasz}.

Note that
the idea of constructing
new symmetries by the repeated commutation of a~master symmetry with a seed
symmetry, as well as the notion of master symmetry, were suggested
by Fuchssteiner and Fokas
\cite{artur:ff}, see also Fuchssteiner \cite{artur:fu}.

Thus, the CDIS equation, as well as the Burgers equation,
represents a nontrivial example of a (1+1)-dimensional evolution equation
possessing a ``doubly infinite"
hereditary algebra of master symmetries $\tau_{m,n}$.

Using (\ref{artur:comm}), it can be shown (cf.\ \cite{artur:blasz}) that
$\mathop{\rm ad}\nolimits_{\tau_{0,j}}^{m+1}(\tau_{m,j'})=0$,
i.e., $\tau_{m,j}$ are master symmetries of degree $m$ for all equations
$u_{t_{k}}=\tau_{0,k}$, $k=0,1,2,\dots$.
Here $\mathop{\rm ad}\nolimits_{B}(G)\equiv [B,G]$ for any
(smooth) functions $B$ and $G$ depending on $\omega$ and on a finite number
of local variables.

Let $\exp(\mathop{\rm ad}\nolimits_{B})\equiv
\sum\limits_{j=0}^{\infty}\mathop{\rm ad}\nolimits^{j}_{B}/j!$.
As $\mathop{\rm ad}\nolimits_{\tau_{0,j}}^{m+1}(\tau_{m,j'})=0$, it
is easy to see (cf.\ \cite{artur:blasz}) that
\begin{gather*}
G_{m,j}^{(k)}(t_k)=\exp(-t_{k}\mathop{\rm ad}
\nolimits_{\tau_{0,k}})\tau_{m,j}\\
\phantom{G_{m,j}^{(k)}(t_k)} =\sum\limits_{i=0}^{m}
 \frac{(-t_{k})^{i}}{i!}
\mathop{\rm ad}\nolimits_{\tau_{0,k}}^{i}(\tau_{m,j})=
\sum\limits_{i=0}^{m}
\frac{((2k+1)t_{k})^{i}m!}{i!(m-i)!} \tau_{m-i,j+ik}
\end{gather*}
are local time-dependent generalized symmetries for the equation
$u_{t_{k}}=\tau_{0,k}$, and $\mathop{\rm ord}\nolimits
G_{m,j}^{(k)}$ $=2(j+mk)+1$. Note that $G_{m,j}^{(k)}$ obey the same
commutation relations as $\tau_{m,j}$, that is,
\begin{equation}
[G_{m,j}^{(k)}, G_{m',j'}^{(k)}]=((2j'+1)m
-(2j+1)m') G_{m+m'-1,j+j'}^{(k)}.\label{artur:comm1}
\end{equation}
It is straightforward to verify that
$\tau_{0,1}=F=u_{3}+3 u^{2} u_{2}+9 u u_{1}^{2}+ 3 u^{4} u_{1}$
and thus
$G_{m,j}=G_{m,j}^{(1)}(t)=\exp(-t\mathop{\rm ad}\nolimits_{F})\tau_{m,j}$
are time-dependent symmetries for the CDIS equation.

It is easy to see that the
number of symmetries $G_{m,j}$ of given odd order $k=2l+1$ equals
$[k/2]+1=l+1$. As $\dim S_{{\rm CDIS},k}\leq [k/2]+1$, these
symmetries exhaust the space $S_{{\rm CDIS},k}$ for odd $k$.
Moreover \cite{snmp}, any local generalized symmetry of the CDIS equation
is a linear combination
of the symmetries $G_{m,j}$ for $m=0,1,2,\dots$ and $j=0,1,2\dots$.

Evaluating the commutator of $W$
with $\tau_{m,0}=x^{m} u_1 + m x^{m-1} u/2$,
we readily see that it vanishes, if we assume that
the result of application of $D^{-1}$ to a homogeneous
polynomial in
$x,t,u,\dots,u_k$ without $u_j$-independent terms is a
polynomial of the same kind. Next, under the same assumption
$[W,\tau_{0,1}]=[W,F]=0$, just because $W\in S_{\mathrm{CDIS}}$.
Finally, using the
commutation relations (\ref{artur:comm})
and the Jacobi identity, we conclude that the
commutator of
$W$ with all symmetries $G_{m,j}$ vanishes as well. As a result, we have

\begin{theorem}\label{artur:t1}
Any symmetry of the CDIS equation of the form $G(x,t,\omega,u,\dots,u_s)$
is a~li\-near combination of the symmetries $G_{m,j}$ for $m,j=0,1,2,\dots$,
and of the symmetry $W=\exp(-2\omega)u$, which commutes with all other
symmetries.
\end{theorem}

This result can be generalized to the symmetries of `higher CDIS
equations' $u_{t_k}=\tau_{0,k}$. Recall~\cite{artur:ibrbook} that
because of existence of infinitely many $x,t$-independent local
generalized symmetries the CDIS equation has a formal symmetry of
infinite rank of the form
$\mathfrak{L}=D+u^2+\sum\limits_{j=1}^{\infty}a_{j}D^{-j}$, where
$a_j$ are some $x,t$-independent local functions. Using Lemma~11
from \cite{artur:sok},  we can show that $\mathfrak{L}$ is a
formal  symmetry of rank at least $\ord \tau_{0,k}+3$ for the
equation $u_{t_k}=\tau_{0,k}$, and hence $u^2=\mathop{\rm
res}\nolimits\ln\mathfrak{L}$ is a  conserved density  for all
these equations: $D_{t_k}(u^2)=D(\sigma_k)$, where $\sigma_k$ are
some local functions. Note that $\sigma_k$ can be choosen to be
polynomials in $u_j$ with $x,t$-independent coefficients and zero
free term (in particular, we have chosen $\sigma_1=2u u_2 + 6 u^3
u_1+u^6-u_1^2=\p\omega/\p t$ in (\ref{om}) to be exactly of this
form).

In order to describe the symmetries of the systems
$u_{t_k}=\tau_{0,k}$, let us extend
(\ref{om})
by adding
the equations
\[
\p\omega/\p t_k=\sigma_k.
\]

\begin{theorem}
Any symmetry of the form $G(x,t_k,\omega,u,\dots,u_s)$ for the equation
$u_{t_k}=\tau_{0,k}$, $k\geq 1$, is a linear combination
of the symmetries $G_{m,j}^{(k)}(t_k)$ for $m=0,1,2,\dots$ and
$j=0,1,2\dots$, and of $W=\exp(-2\omega)u$, which commutes with all
other symmetries.
\end{theorem}

As a final remark, let us mention that, as an alternative approach, one
could try using the formula $v=\exp(\omega) u$
in order to ``transfer'' the known results on symmetries and recursion operators of
the linear equation $v_t=v_3$ to the CDIS equation.
However, the symmetries
resulting in this way are in general highly nonlocal. Even the description of
nonlocalities that occur in thus constructed symmetries, as
well as sorting out local and ``weakly nonlocal'' (depending only on $\omega$)
symmetries, is quite a difficult task, and we intend to analyze this and related
problems elsewhere.

\subsection*{Acknowledgements}

The research of AS was supported by DFG
via Graduierten\-kolleg ``Geometrie
und Nicht\-lineare Analysis'' at Humboldt-Universit\"at zu Berlin, Germany,
where he held a postdoctoral fellowship, as well as by the
Ministry of Education, Youth and Sports of Czech Republic under Grant
MSM:J10/98:192400002,  and by Grant No.201/00/0724 from the Czech Grant
Agency. AS acknowledges with gratitude the hospitality of Faculty of
Mathematics and Computer Science of the Vrije Universiteit Amsterdam
during his two-week visit when the present work was initiated.

The authors thank Jing Ping Wang for her comments.
It is also our pleasure to thank the
referees for valuable comments and suggestions.

\label{artur-lastpage}

\end{document}